\documentclass[reprint,prb,amsmath,amssymb]{revtex4-1}
\usepackage{graphicx}
\usepackage{dcolumn}
\usepackage{bm}

\begin{document}
\title{Tracking electron pathways with magnetic field: Aperiodic Aharonov-Bohm oscillations in coherent transport through a periodic array of quantum dots}
\author{L. S. Petrosyan,$^{1,2}$ and T. V. Shahbazyan$^1$} 
\affiliation{
$^1$Department of Physics, Jackson State University, Jackson, Mississippi
39217 USA\\
$^2$Department of Physics, Russian-Armenian University, 123  Emin Street, Yerevan, 0021 Armenia
}
\begin{abstract}
We study resonant tunneling through a periodic square array of quantum dots sandwiched between modulation-doped quantum wells. If a magnetic field is applied parallel to the quantum dot plane, the tunneling current exhibits a highly complex Aharonov-Bohm oscillation pattern due to the interference of multiple pathways traversed by a tunneling electron. Individual pathways associated with conductance beats can be enumerated by sweeping the magnetic field at various tilt angles. Remarkably, Aharonov-Bohm  oscillations are \textit{aperiodic} unless the magnetic field slope relative to the quantum dot lattice axes is a rational number. 
\end{abstract}
\maketitle

\section{Introduction}
\label{intro}

Interference effects in quantum transport in semiconductor quantum dots (QD) have been among the highlights in electron  transport studies.\cite{yacoby-prl95,shuster-nature97,buks-nature98} A simple example of coherent transport is resonant tunneling through a pair of QDs independently coupled to left and right doped semiconductor leads that shows a conductance peak narrowing due to the interference between the tunneling electron pathways.\cite{shahbazyan-prb94}  In the presence of magnetic field, the tunneling current through a QD system exhibits Aharonov-Bohm (AB) oscillations \cite{blick-prl01,ensslin-prl06,hatano-prl11} as a function of magnetic flux through a surface enclosed  by pathways.\cite{shahbazyan-prb94,loss-prl00,gefen-prl01} AB oscillations have been widely studied in systems where electron motion is constrained by the system geometry such as, e.g., metal or semiconductor rings,\cite{gefen-prl84,webb-prl85,chandrasekhar-prl85,benoit-prl97} carbon nanotubes \cite{ando-pb94,bachtold-nature99,kono-science04} or, more recently, graphene nanorings.\cite{russo-prb08,ensslin-njp10} 

At the same time, in \textit{open} two-dimensional (2D) electron systems, i.e., when the electron motion in a 2D plane is unconstrained, oscillations of magnetoresistance were observed in the presence of a weak 1D \cite{wei89,win89,bet90} or 2D \cite{alv89,wei90,fan90} periodic potential and in a system of antidots.\cite{ens90,wei91,lor91,wei93} In such structures, the oscillations are caused by geometric resonances occurring when the size of the electron's Larmor orbit, which changes with magnetic field, is commensurate with the potential period.\cite{ger89} For magnetic fields corresponding to  magnetoresistance maxima, electron trajectories run close to the potential energy minima, indicating that oscillations originate from electron orbital motion rather than its phase.

\begin{figure}[tb]
\begin{center}
\includegraphics[width=0.85\columnwidth]{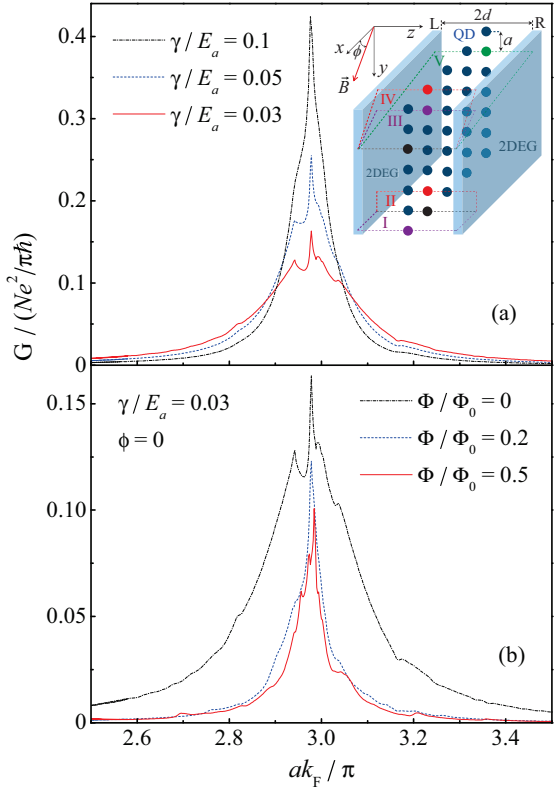}
\caption{\label{fig:1}
(Color online) Normalized conductance vs. electron Fermi momentum. (a) Emergence of sidebands for zero-field conductance with decreasing 2DEG scattering rate. (b) Conductance lineshape change with varying in-plane magnetic field. Inset: Schematic view of a QD lattice placed between 2DEGs with several closed electron pathways.}
\end{center}
\end{figure}

Here we show that AB oscillations can occur in an open 2D system where electron transport takes place via multiple pathways. We consider resonant tunneling through a square periodic array of QDs  sandwiched between two 2D electron gases (2DEGs)  in doped semiconductor quantum wells  separated from the QD plane by tunneling barriers (see the inset in Fig.\ \ref{fig:1}). Highly periodic square arrays of QDs have been recently manufactured.\cite{bandyopadhyay-nano96,liang-apl04,zin-nano09} Tunneling current, e.g.,  from left to right 2DEGs involves an electron traversing back and forth along closed pathways comprising electron trajectories within 2DEGs and tunneling between them through QD lattice sites (we assume that direct interdot coupling is negligibly small).  For each closed pathway, an in-plane magnetic field ${\bm B}$ generates a flux $BS_{n}$, where $S_{n}$ is the area of the surface enclosed by such a pathway projected onto the plane normal to ${\bm B}$. We demonstrate that the array magnetoconductance  exhibits a highly complex AB oscillation pattern originating from multiple pathways traversed by the tunneling electron (see the inset in Fig.\ \ref{fig:1}). For high mobility 2DEG characterized by a mean free path $l$ that is much larger than the QD lattice constant $a$, the conductance AB beats correspond to pathways of length $L\lesssim l$ which, hence, can be tracked by sweeping the magnetic field. Remarkably, conductance oscillations are \textit{aperiodic} unless the magnetic field slope relative to the QD lattice axes is a rational number. The lack of AB beat periodicity for a general field orientation implies the absence of pathway degeneracies caused by two or more pathways accommodating the same flux.
%


\section{Conductance through a periodic array of quantum dots}
\label{theory}

To obtain electron  conductance through a QD array, we adopt the tunneling Hamiltonian formalism.\cite{petrosyan-prl11}  The Hamiltonian of a square lattice of $N$ QDs with in-plane coordinates ${\bm r}_{j}$ separated by potential barriers from left and right 2DEG planes has the form
\begin{equation}
\label{h}
H=\sum_{j} E_{0}c_{j}^{\dagger}c_{j}+ 
\sum_{{\bm k}\alpha}{\cal E}_{{\bm k}}^{\alpha}c_{{\bm k}\alpha}^{\dagger}c_{{\bm k}\alpha}+
\sum_{\nu\alpha j} \left (V_{j{\bm k}}^{\alpha} c_{j}^{\dagger}c_{{\bm k}\alpha}+\text{H.c.}\right ),
\end{equation}
where  $c_{j}^{\dagger}$, $c_{j}$, and $ E_{0}$ are, respectively, the creation and annihilation operators and energies for QD localized states, $c_{{\bm k}\alpha}^{\dagger}$, $c_{{\bm k}\alpha}$, and ${\cal E}_{{\bm k}}^{\alpha}$ are those for 2DEG states ($\alpha=L,R$), and $V_{{\bm k} j}^{\alpha}$ is a transition matrix element between localized and 2DEG states. We assume that direct tunneling between QDs is weak and does not include interdot coupling in Hamiltonian (\ref{h}). We restrict ourselves to the single-electron picture of transport and disregard electron interaction effects due to a low probability of QD double occupancy in a large array. We also assume that the magnetic field, to be included below, is sufficiently weak and/or the electron \textit{g}-factor is sufficiently small  to cause significant Zeeman splitting and suppress spin indices throughout. The zero-temperature conductance through a system of $N$ QDs is given by \cite{brandes-pr05}
\begin{equation}
\label{cond}
G=\frac{e^{2}}{\pi\hbar}\, {\rm Tr} \left(\hat{\Gamma}^{R}\frac{1}{E_{F}-E_{0}-\hat{\Sigma}} \,\hat{\Gamma}^{L}\frac{1}{E_{F}-E_{0}-\hat{\Sigma}^{\dagger}}\right),
\end{equation}
where $\Sigma_{ij}=\Sigma_{ij}^{L}+\Sigma_{ij}^{R}$ is the self-energy matrix of QD states due to coupling to electron states in the left and right 2DEGs, 
\begin{equation}
\label{self-matrix}
\Sigma_{ij}^{\alpha} =\sum_{ \nu}\frac{V_{i{\bm k}}^{\alpha}V_{{\bm k} j}^{\alpha}}{E_{F} -{\cal E}_{{\bm k}}^{\alpha}+i\gamma_{\alpha}} = \Delta_{ij}^{\alpha}-\frac{i}{2} \Gamma_{ij}^{\alpha}.
\end{equation}
Here the principal and singular parts of $\Sigma_{ij}^{\alpha}$ determine the energy matrix $\Delta_{ij}^{\alpha}$ and the decay matrix $\Gamma_{ij}^{\alpha}$, respectively, and the trace is taken over QD lattice sites. The transition matrix element can be presented as \cite{shahbazyan-prb94} $V_{j{\bm k}}^{\alpha}=A^{-1/2}e^{i{\bm k}\cdot {\bm r}_{j}}t_{\alpha}$, where ${\bm k}$ and ${\bm r}_{j}$ are, respectively, the electron momentum and coordinate in 2DEGs, $t_{\alpha}$ is  the tunneling amplitude between QD and 2DEG, and $A=Na^{2}$ is the normalization area. We assumed that the barrier is sufficiently high so that electron tunneling between 2DEG and the QD plane takes place along the shortest path and the dependence of $t_{\alpha}$ on energy is weak.\cite{shahbazyan-prb94}  Then the self-energy (\ref{self-matrix}) takes the form $\Sigma_{ij}^{\alpha} =t_{\alpha}^{2}G_{\alpha}({\bm r}_{i}-{\bm r}_{j})$, where $G_{\alpha}({\bm r}_{i}-{\bm r}_{j})$ is the electron Green's function between the QD lattice sites projected onto 2DEG planes.

The coupling between the QD lattice states and the continuum of electronic states in 2DEGs gives rise to an \textit{in-plane} quasimomentum ${\bm p}$ that conserves across the system \cite{petrosyan-prl11}. The 2DEG momentum space ${\bm k}$ splits into Bloch bands ${\bm k}={\bm g}+{\bm p}$, where ${\bm g}=(2\pi m/a,2\pi n/a)$ are reciprocal lattice vectors ($m$ and $n$ are integers) and  ${\bm p}$ lies in the first 2D Brillouin zone ($-\pi/a <p_{x},p_{y}< \pi/a$). The energy spectrum of the QD lattice states can be obtained by performing  a Fourier transform of the self-energy matrix Eq.~(\ref{self-matrix}) as  $\Sigma_{ij}^{\alpha}=N^{-1}\sum_{\bm p}e^{i{\bm p}\cdot ({\bm r}_{j}-{\bm r}_{j})}\Sigma_{\bm p}^{\alpha}$, where
%
\begin{equation}
\label{self-lsb}
\Sigma_{\bm p}^{\alpha}=\frac{t_{\alpha}^{2}}{a^{2}}\sum_{\bm g}G^{\alpha}_{{\bm p}+{\bm g}} =\frac{t_{\alpha}^{2}}{a^{2}}\sum_{\bm g} \frac{1}{E_{F} -{\cal E}_{{\bm g}+{\bm p}}^{\alpha}+i\gamma_{\alpha}}.
\end{equation}
Here $G^{\alpha}_{{\bm g}+{\bm p}}$ is the momentum space 2DEG Green's function of the band ${\bm g}$  electron having a quasimomentum ${\bm p}$, ${\cal E}_{{\bm g}+{\bm p}}^{\alpha}=\hbar^{2}\left ({\bm g}+{\bm p}\right )^{2}/2m_{\alpha}$ is its dispersion ($m_{\alpha}$ is the electron mass), and $\gamma_{\alpha}$ is its scattering rate. In momentum space, the self-energy $\Sigma_{\bm p}^{\alpha}=\Delta_{\bm p}^{\alpha}-\frac{i}{2}\Gamma_{\bm p}^{\alpha}$ is a complex function of ${\bm p}$ that determines the QD lattice band dispersion, $E_{\bm p}=E_{0}+\Delta_{\bm p}^{L}+\Delta_{\bm p}^{R}$, and its decay width, $\Gamma_{\bm p}^{L}+\Gamma_{\bm p}^{R}$, due to the coupling to left and right 2DEGs.

We now include an in-plane magnetic field ${\bm B}$ tilted by an angle $\phi$ relative to the $x$-axis (see the inset in Fig.\ \ref{fig:1}) through the vector potential ${\bm A}=(B_{y}z, -B_{x}z,0)$. This leads to a momentum shift in the left and right 2DEGs, located at $z=-d$ and $z=d$, respectively, as ${\bm k} + (e/\hbar c) {\bm A}^{\alpha}$, where  ${\bm A}^{L,R}=\mp dB(\sin\phi, -\cos\phi)$. In the presence of the QD lattice, the momentum space in the left or right 2DEG is now split as ${\bm k}^{\alpha} = {\bm g}^{\alpha}_{B}+{\bm p}$, where
\begin{equation}
\label{band}
{\bm g}_{B}^{L,R}=\frac{2\pi}{a}\left (m\mp \frac{\Phi}{2\Phi_{0}}\sin\phi, n\pm \frac{\Phi}{2\Phi_{0}}\cos\phi\right )
\end{equation}
is a field-dependent band wave vector. Here $\Phi=2daB$ is the magnetic flux through the {\em elementary area} enclosed by pathways running between 2DEGs (pathways I or II in the inset of Fig.\ \ref{fig:1}) and $\Phi_{0}=hc/e$ is the flux quantum. The field dependence of ${\bm g}_{B}^{L,R}$ in the 2DEG electron dispersion ${\cal E}_{{\bm g}^{\alpha}_{B}+{\bm p}}^{\alpha}$ translates to field-dependence of the QD lattice self-energy, $\Sigma_{\bm p}^{\alpha}(B)=\Delta_{\bm p}^{\alpha}(B)-\frac{i}{2}\Gamma_{\bm p}^{\alpha}(B)$, still given by Eq.\ (\ref{self-lsb}) but with ${\bm g}$ replaced by ${\bm g}^{\alpha}_{B}$. Finally, the array conductance is obtained via a Fourier transform of Eq.~(\ref{cond}) as 
\begin{equation}
\label{cond2}
G=\dfrac{N e^{2}}{\pi\hbar}\,a^{2}\! \! \! \int \! \! \! \frac{d{\bm p}}{(2\pi)^{2}} \,
\frac{\Gamma_{\bm p}^{L}\Gamma_{\bm p}^{R}}{\left (E_{F}-E_{\bm p}\right ) ^{2}+\frac{1}{4}\left (\Gamma_{\bm p}^{L}+\Gamma_{\bm p}^{R}\right )^{2}},
\end{equation}
where the ${\bm p}$ integral is taken over the 2D Brillouin zone.  The field dependence of $G(B)$ comes from those of the QD lattice band dispersion $E_{\bm p}(B)=E_{0}+\Delta_{\bm p}^{L}(B)+\Delta_{\bm p}^{R}(B)$ and its width $\Gamma_{\bm p}^{\alpha}(B)$.

\section{Discussion and numerical results}
\label{num}

Changing the magnetic field magnitude may cause a 2DEG electron to jump to another band, according to Eq.\ (\ref{band}). Since the QD lattice energy spectrum, $E_{\bm p}(B)-i\Gamma_{\bm p}^{\alpha}(B)/2$, includes contributions from all 2DEG bands, these field-induced interband transitions lead to oscillatory behavior of $E_{\bm p}(B)$ and $\Gamma_{\bm p}^{\alpha}(B)$ which, in turn, gives rise to AB conductance oscillations. Importantly, depending on the field orientation $\phi$, interband transitions for the $x$ and $y$ components of ${\bm g}_{B}^{\alpha}$ can take place at \emph{different} field values. For example, for ${\bm B}$ oriented along the $x$ or $y$ axes ($\phi=0$ or $\pi/2$), only one component of  ${\bm g}_{B}^{\alpha}$ can jump to the next value ($n\rightarrow n\pm 1$ or $m\rightarrow m\mp 1$) with changing field magnitude [see Eq.\ (\ref{band})]; in real space, only the surfaces' projection onto the $(yz)$ or $(xz)$ planes, respectively, would contribute to oscillations (e.g., either pathways II or I in the inset of Fig.\ \ref{fig:1}).  However, for $\phi=\pi/4$, interband transitions  simultaneously take place  for both components of ${\bm g}_{B}^{\alpha}$ ($n\rightarrow n\pm 1$ and $m\rightarrow m\mp 1$), i.e.,  pathways I and II would lead to similar oscillations. Note that an interband transition involves contributions from many pathways, but, due to the square lattice symmetry, only two independent sets of oscillations in 2D momentum space, described by Eq.\ (\ref{band}), are generated by all electron pathways. For a general field orientation, these two oscillation sets are incommensurate, implying that the resulting AB oscillation pattern is \emph{aperiodic}. The AB beats are periodic for field orientations that render commensurate interband transitions for both ${\bm g}_{B}^{\alpha}$ components, i.e., for $\tan \phi = p/q$, where $p$ and $q$ are integers. In other words, the fluxes through projections of a surface, enclosed by an electron pathway, onto the $(yz)$ and $(xz)$ planes must be commensurate, which only takes place, given the lattice symmetry, if the magnetic field slope is a rational number.

Below we present numerical calculations for a symmetric configuration, i.e., for a QD lattice at midpoint between similar quantum wells ($\gamma_{\alpha}=\gamma$ and $m_{\alpha}=m$, $t_{\alpha}=t$). The lattice constant was chosen to set $E_{0}=9E_{a}$, where $E_{a}=\pi^{2}\hbar^{2}/2ma^{2}$ is a geometric energy scale associated with the lattice, so that the transmission resonance $E_{F}=E_{0}$ occurs at a Fermi momentum $k_{F}=3\pi/a$. The 2DEG electron width $\gamma=\hbar v_{F}/l$ due to elastic scattering was varied in the range from  $0.01E_{a}$ to $0.1E_{a}$, yielding $l/a$ in the range from 20 to 200; for the lattice period $a=20$ nm, this corresponds to a low-to-intermediate 2DEG mobility in the range  $10^{4}$--$10^{6}$ cm$^{2}$/Vs.

In Fig.~\ref{fig:1}(a), zero-field per QD normalized conductance is shown for several values of $\gamma$. For low mobility 2DEG with $\gamma/E_{a}=0.1$, the conductance shows a single peak centered at a QD resonance with weak shoulders on the left and right sides. With decreasing  $\gamma$, these shoulders develop into sidebands while the main peak gets slightly shifted. These features are due to the appearance of new resonances in the integrand of Eq.\ (\ref{cond2}) satisfying $E_{F}=E_{\bm p}$ originating from QD lattice coupling to 2DEGs [see Eq.\ (\ref{self-lsb})]. At the same time, with decreasing $\gamma$, the electron escape rate from the QD lattice to 2DEG, $\Gamma_{\bm p}=-2{\rm Im}\Sigma_{\bm p}$, becomes a sharp function of $E_{F}$. The combination of these two factors results in sharp features  near the resonance and in the emergence of minor features away from it, the latter coming from neighboring bands. 

\begin{figure}[tb]
\begin{center}
\includegraphics[width=0.85\columnwidth]{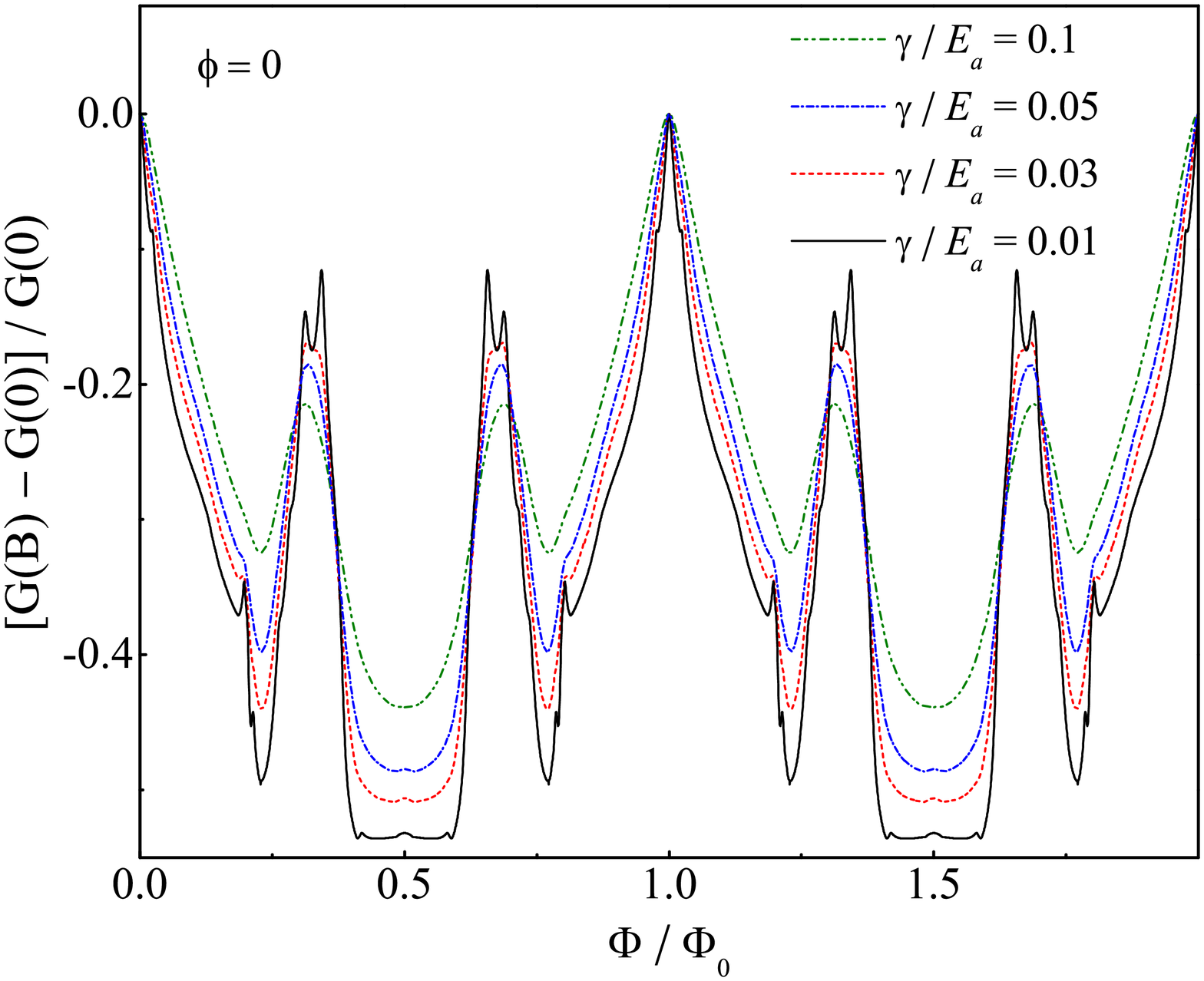}
\caption{\label{fig:2}
(Color online) AB oscillations of magnetoconductance for the 2DEG scattering rate $\gamma$ in the range $0.1E_{a}$--$0.01E_{a}$. With increasing 2DEG mobility, the oscillation pattern develops a fine structure associated with longer electron pathways.}
\end{center}
\end{figure}
\begin{figure}[tb]
\begin{center}
\includegraphics[width=0.85\columnwidth]{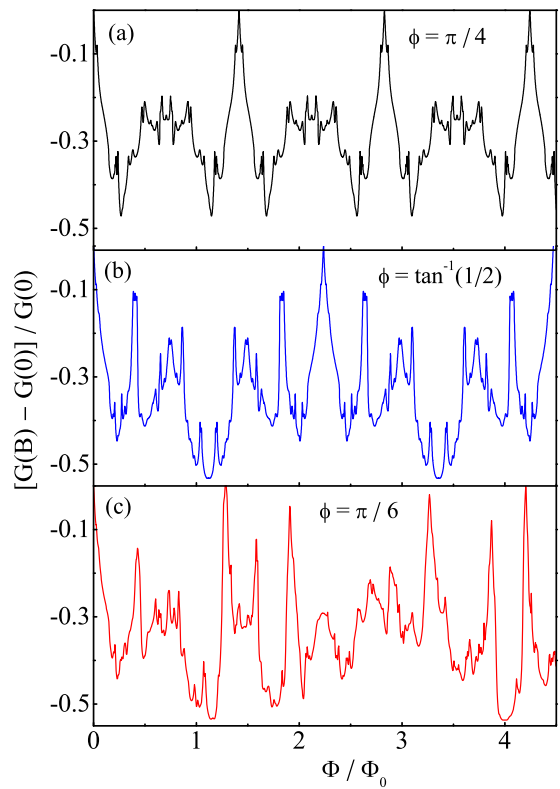}
\caption{\label{fig:3}
(Color online) AB oscillations of magnetoconductance at $\gamma/E_{a}=0.01$ for a magnetic field slope equal (a) 1,  (b) 1/2, and (c)  $3^{-1/2}$ . The oscillation pattern in (c) is aperiodic.}
\end{center}
\end{figure}

An in-plane magnetic field leads to significant decrease of the overall conductance and to a change of sidebands' positions and widths [see Fig.~\ref{fig:1}(b)]. While the latter behavior reflects the  field dependence of $E_{F}=E_{\bm p}(B)$ resonances, the amplitude drop comes from the change of interference between tunneling paths caused by AB flux.  Sweeping the magnetic field reveals  pronounced AB oscillations of peak conductance with amplitude exceeding half of its zero-field value (see Fig.\ \ref{fig:2}). For $\phi =0$, the largest period in AB oscillations pattern, in units of flux through elementary area $S_{0}=2da$, is provided by pathways enclosing surfaces that project area $S_{0}$ onto the $(yz)$ plane, e.g., pathways II, III, and V in Fig.\ \ref{fig:1} inset, while strong half period beats come from pathways enclosing area $2S_{0}$ when projected onto $(yz)$ plane, e.g., pathway IV. With decreasing $\gamma$, conductance oscillations develop \emph{fine structure} due to the short-period beats coming from longer pathways. 

Tilting the magnetic field reveals a dramatic increase of the fine structure complexity due to the  reduction of pathways degeneracies, which are maximal for ${\bm B}$ oriented along the lattice axes (see Fig.\ \ref{fig:3}). Individual beats in the AB oscillation pattern correspond to specific pathways traversed by an electron, i.e., a varying magnetic field enumerates the pathways by highlighting those with surfaces accommodating the integer AB flux (in units of $\Phi_0$). Note that the pathway degeneracies persist for any rational field slope (i.e., $\tan \phi =p/q$); e.g., for $\tan \phi =1$, pathways I and II accommodate the same flux $\Phi/\sqrt{2}$ corresponding to the largest period in Fig.\ \ref{fig:3}(a). By changing the field slope,  a substantially different oscillation pattern is generated that highlights a different set of pathways. For example, for $\tan \phi =1/2$ [see Fig.\ \ref{fig:3}(b)], pathways I and II are now distinct by producing beats with periods 2.24 and 1.12, respectively (in units of $\Phi$). Note also that, for any rational slope, there are "missing" pathways that do not produce AB beats, e.g., pathway I for $\phi=0$, II for $\phi=\pi/2$,  III for $\phi=\pi/4$, IV for $\tan \phi = 2$, and V for $\tan \phi = 1/4$. For a general field slope, however, the AB oscillation pattern has no periodic structure and there are no degenerate or missing pathways. An example of aperiodic beats for $\tan\phi = 3^{-1/2}$ is shown in Fig.\ \ref{fig:3}(c).

\section{Conclusion}
\label{conc}

In summary, we have shown that the tunneling current through a periodic array of quantum dots sandwiched between 2D electron gases in quantum wells exhibits a highly complex pattern of Aharonov-Bohm oscillations originating from multiple pathways that the electron traverses in the course of transport. For high mobility samples, the AB beats corresponding to individual pathways are well resolved and could allow tracking the electron motion in the system by sweeping the magnetic field. We find that oscillation pattern is aperiodic unless the magnetic field slope relative to the lattice axes is a rational number.

\acknowledgments

This work was supported by the National Science Foundation under Grant No. DMR-1206975 and CREST center.


\begin{thebibliography}{}

\bibitem{yacoby-prl95} A. Yacoby, M. Heiblum, D. Mahalu, and H. Shtrikman, Phys. Rev. Lett. \textbf{74}, 4047 (1995).

\bibitem{shuster-nature97} R. Schuster, E. Buks, M. Heiblum, D. Mahalu, V. Umansky, and H. Shtrikman, 
Nature (London) \textbf{385}, 417 (1997).

\bibitem{buks-nature98} E. Buks, R. Schuster, M. Heiblum, D. Mahalu, and V. Umansky, 
Nature (London) \textbf{391}, 871 (1998).

\bibitem{shahbazyan-prb94}T. V. Shahbazyan and M. E. Raikh, Phys. Rev. B 49, 17123 (1994).

\bibitem{blick-prl01} A. W. Holleitner, C. R. Decker, H. Qin, K. Eberl, and R. H. Blick, 
Phys. Rev. Lett. \textbf{87}, 256802 (2001).

\bibitem{ensslin-prl06} M. Sigrist, T. Ihn, K. Ensslin, D. Loss, M. Reinwald, and W. Wegscheider, 
Phys. Rev. Lett. \textbf{96}, 036804 (2006).

\bibitem{hatano-prl11} T. Hatano, T. Kubo, Y. Tokura, S. Amaha, S. Teraoka, and S. Tarucha, 
Phys. Rev. Lett. \textbf{106}, 076801 (2011). 

\bibitem{loss-prl00} D. Loss and E. V. Sukhorukov, Phys. Rev. Lett. \textbf{84}, 1035 (2000).

\bibitem{gefen-prl01} J. K\"{o}nig and Y. Gefen, Phys. Rev. Lett. \textbf{86}, 3855 (2001).

\bibitem{gefen-prl84} Y. Gefen, Y. Imry, and M. Ya. Azbel,
Phys. Rev. Lett. \textbf{52}, 129 (1984).

\bibitem{webb-prl85} R.A. Webb, S. Washburn, C.P. Umbach, and R.B. Laibowitz, Phys. Rev. Lett. \textbf{54}, 2696 (1985).

\bibitem{chandrasekhar-prl85}V. Chandrasekhar, M. J. Rooks, S. Wind,  and D. E. Prober, Phys. Rev. Lett. \textbf{55}, 1610 (1985). 

\bibitem{benoit-prl97} G. Cernicchiaro, T. Martin, K. Hasselbach, D. Mailly, and A. Benoit,
Phys. Rev. Lett. \textbf{79}, 273 (1997).

\bibitem{ando-pb94}H. Ajiki, T. Ando, Physica B \textbf{201}, 349 (1994).

\bibitem{bachtold-nature99}A. Bachtold, C. Strunk, J.-P. Salvetat, J.-M. Bonard, L. Forr\'{o}, T. Nussbaumer, and C. Sch\"{o}nenberger,
Nature (London) \textbf{397}, 673 (1999).

\bibitem{kono-science04}S. Zaric, G. N. Ostojic, J. Kono, J. Shaver, V. C. Moore, M. S. Strano, R. H. Hauge, R. E. Smalley, and X. Wei, 
Science \textbf{304}, 1129 (2004).

\bibitem{russo-prb08} S. Russo, J. B. Oostinga, D. Wehenkel, H. B. Heersche, S. S. Sobhani, L. M. K. Vandersypen, and A. F. Morpurgo,
Phys. Rev. B \textbf{77}, 085413 (2008).

\bibitem{ensslin-njp10}M. Huefner, F. Molitor, A. Jacobsen, A. Pioda, C. Stampfer, K. Ensslin and T. Ihn,
New J. Phys. \textbf{12} 043054 (2010).

\bibitem{wei89}D. Weiss, K. v. Klitzing, K. Ploog, and G. Weinmann, 
Europhys.\ Lett.\ {\bf 8}, 179 (1989).

\bibitem{win89}R. W. Winkler, J. P. Kotthaus, and K. Ploog, 
Phys.\ Rev.\ Lett.\ {\bf 62}, 1177 (1989).

\bibitem{bet90}P. H. Beton, E. S. Alves, P. C. Main, L. Eaves, M. W. Dellow, 
M. Henini, O. H. Hughes, S. P. Beaumont, and C. D. W. Wilkinson, 
Phys. Rev.~B {\bf 42}, 9229 (1990).

\bibitem{alv89} E. S. Alves, P. H. Beton, M. Henini, L. Eaves, P. C. Main, 
O. H. Hughes, G. A. Toombs, S. P. Beaumont, and C. D. W. Wilkinson, 
J.\ Phys.\ Condens.\ Matter, {\bf 1}, 8257 (1989).

\bibitem{wei90}D. Weiss, K. v. Klitzing, K. Ploog, and G. Weimann, 
Surf.\ Sci.\ {\bf 229}, 88 (1990).

\bibitem{fan90}H. Fang and P. J. Stiles, 
Phys. Rev.~B {\bf 41}, 10171 (1990).

\bibitem{ger91}R. R. Gerhardts, D. Weiss, and U. Wulf, 
Phys. Rev.~B {\bf 43}, 5192 (1991).


\bibitem{ens90}K. Ensslin and P. M. Petroff, 
Phys. Rev.~B {\bf 41}, 12307 (1990).

\bibitem{wei91}D. Weiss, M. L. Roukes, A. Menschig, P. Grambow, 
K. von Klitzing, and G. Weimann, 
Phys.\ Rev.\ Lett.\ {\bf 66}, 2790 (1991).

\bibitem{lor91}A. Lorke, J. P. Kotthaus, and K. Ploog, 
Phys.\ Rev.\ {\bf 44}, 3447 (1991).


\bibitem{wei93}D. Weiss, K. Richter, A. Menschig, R. Bergmann, 
H. Schweizer, K. von Klitzing, and G. Weimann, 
Phys.\ Rev.\ Lett.\ {\bf 70}, 4118 (1993).


\bibitem{ger89}R. R. Gerhardts, D. Weiss, and K. von Klitzing, 
Phys. Rev. Lett.\ {\bf 62}, 1173 (1989).


\bibitem{bandyopadhyay-nano96} S. Bandyopadhyay, A. E. Miller, H. C. Chang, G. Banerjee, V. Yuzhakov, D.-F. Yue, R. E. Ricker, S. Jones, J. A. Eastman, E. Baugher, and M. Chandrasekhar, 
Nanotechnology \textbf{7}  360 (1996).


\bibitem{liang-apl04} J. Liang, H. Luo, R. Beresford, and J. Xu,
Appl. Phys. Lett. \textbf{85}, 5974 (2004).

\bibitem{zin-nano09} M. T. Zin, K. Leong, N.-Y. Wong, H. Ma,
M. Sarikaya, and A. K.-Y. Jen, 
Nanotechnology \textbf{20},  015305 (2009).

\bibitem{petrosyan-prl11} L. S. Petrosyan, A. S. Kirakosyan, and T. V. Shahbazyan, 
Phys. Rev. Lett.\ \textbf{107}, 196802 (2011).

\bibitem{brandes-pr05} T. Brandes, Phys. Rep. \textbf{408}, 315 (2005).

\end{thebibliography}
\end{document}